# Proposition of a full deterministic medium access method for wireless network in a robotic application

Adrien van den Bossche, Thierry Val, Eric Campo
ICARE Research Team
Blagnac, France
{vandenbo, val, campo}@iut-blagnac.fr

*Abstract*—Today, many network applications require shorter react time. Robotic field is an excellent example of these needs: robot react time has a direct effect on its task's complexity. Here, we propose a full deterministic medium access method for a wireless robotic application. This contribution is based on some low-power wireless personal area networks, like *ZigBee* standard. Indeed, *ZigBee* has identified limits with Quality of Service due to non-determinist medium access and probable collisions during medium reservation requests. In this paper, two major improvements are proposed: an efficient polling of the star nodes and a temporal deterministic distribution of peer-to-peer messages. This new MAC protocol with no collision offers some QoS faculties.

## I. INTRODUCTION

Today, networks and telecoms are widely used. In production, space technologies, or even at home, networks and moreover wireless networks are considered to become essential. Wireless technologies eliminate expensive, heavy, not aesthetic cables, which are not easy to install or to use. Using a wireless connection is sometimes a luxury, but it may be necessary in many cases of moving devices, like car tire sensors, etc. All these points encourage research and industrial to develop technologies and products in this domain.

Nowadays in telecoms and networks fields, developers are essentially trying to make the products cheaper and easier to use. Baudrate is constantly increasing and there is currently a high level of performance and compatibility between technologies which are globally mature. Unfortunately, there is a technical problem which is not yet solved: temporal performance on the delivery of the network messages. In the 80's, Internet network applications did not need good temporal performance because there were no real-time applications over the network. Today's applications require higher temporal performances: audio/video streaming for home applications, sensors networks [1], medical assistance [2], etc. are widely used and imply temporal certitudes; medium access needs to *be determinist* and a maximum latency has to be kept within identified limits. In many networks, errors are generally corrected by protocols with an acknowledge/retransmit mechanism; for many real-time applications, this method can't be used because of the latency introduced by retransmissions [3].

Another point that can be noticed in this introduction is the interest to regroup network and embedded sciences. Today's wireless devices are portable; because of it, energy saving becomes a major goal in the development of an electronic device. This point is crucial and developers have to ask themselves about conventional methods for solving classical problems. For example unnecessary retransmissions, padding, frame collisions or permanent listening of an unused medium are energy losses which must be limited. IEEE recently worked on Low Power Wireless Personal Area Network (LP-WPAN) and some new technologies like IEEE 802.15.4/ZigBee can solve many problems on simple communicating embedded devices. This work plans to improve network protocols for optimal energy saving.

## II. OUR WIRELESS ROBOTIC APPLICATION

### A. General presentation and objectives

Our application takes place in mobile robotic field. Mobile robots are obviously embedded devices. They can communicate with a central entity of command which can be mobile/fixed or with other robots. In that case, robots are *cooperating*. It has two goals: first, eliminating many wires in mobile robots which are equipped with sensors. Research prototypes robots will be easier to repair and to upgrade. Moreover, robots will be lighter which can be fundamental for unearthed vehicles. The other goal of the application is the creation of communication channels between robots by this internal network. By this way, cooperating robots can listen to messages of other robots and use this listening for learning about their environment. This capacity is interesting for cooperating robots: regulate pursuit speed, mapping environment, etc. For example, if one of the robots breaks down because of an external physical problem (due to a fall, excessive heat, clash, etc.), others robots can intercept the message of the sensor and react to help or protect themselves.

### B. Network proposed

As an illustration of this work, our application plans to create two types of communication channels as illustrated in fig. 1: between a group of sensors/actuators and the robot command unit, and between robot command units.

For the first type of communication, the network will naturally be configured as a *star topology* with a unique star in each mobile robot. Each star is composed of a *coordinator* located in the command unit of the robot and of network nodes for each sensor. Each node establishes communications with the coordinator via *star links* and cannot communicate directly with another node without going through the coordinator.

Mobile robot, i.e. command device, can also communicate with other mobile devices or fixed network infrastructure by establishing *peer-to-peer network links*. Only coordinators can create peer-to-peer links, nodes cannot. The application may also support coordinators without nodes, if slave nodes were all turned off or if there was no node present for this coordinator in the area.

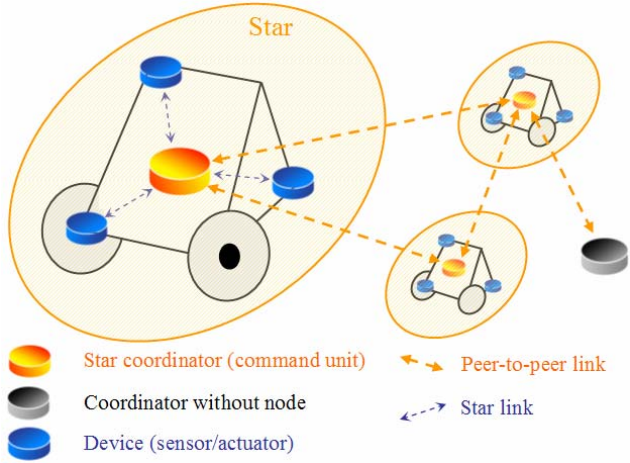

Figure 1.  Network topology inside and between robots

To test this network, we recently designed several network adapters based on the *Freescale semiconductor MC13192* chip [4]. These devices are placed on some ER1 robots, by *Evolution Robotics* [5].

### III. OVERVIEW OF IEEE 802.15.4 STANDARD

In the last years, many IEEE's works study *Low Power Wireless Personal Area Networks* (LP-WPAN) such as IEEE 802.15.4 [6]. This standard proposes an original two-layer protocol stack (*physical-layer* and *data link-layer*) for low power transceivers and low baudrate communications between embedded devices. Energy saving has been optimized by using innovative concepts.

#### A. Nodes types

An IEEE 802.15.4 network is built with several types of devices: *simple nodes*, *routers*, *star coordinators* or *network coordinators*.

802.15.4 standard describes two versions of the protocol stack. On the one hand there is *a Full Function Device* (FFD) stack which includes all functionalities proposed by the standard. A FFD can achieve all functionalities of the network: terminal node, routing and coordinating. On the other hand there is a lighter version of the stack named *Reduced Function Device* (RFD). RFDs can be only a terminal node of the network. There are few levels of reduction of the stack: with or without beacon mode (§III.C.2), with or without crypt functions, etc. Generally, FFDs are main powered and RFDs are embedded.

#### B. Physical layer

IEEE 802.15.4 physical layer has been designed for maximum energy saving:

- DSSS (Direct Sequence Spread Spectrum) O-QPSK (Orthogonal-Quadrature Phase Shift Keying) modulation: excellent noise immunity so transceiver uses low power over the air,
- good receiver selectivity (-95dBm),
- low baudrate: a high noise immunity, few transmission errors so few treatments for higher levels (link and network); using simple processors (8bit) is possible.

All 802.15.4 nodes use the same radio channel. Protocols are optimized for short and periodical data transfers: nodes are sleeping most of the time. In this IEEE standard, this mode is called *doze mode*. It allows ultra low power consumption (40µA) and nodes can become operational in a very short time (330µs) [7]. In doze mode, all radio functionalities are powered down, so it is not possible to receive messages over the air. Dozing devices must establish together a future waking moment before switching doze mode (synchronous wake up), but sleeping devices can also wake up on a local event (asynchronous wake up), for example on a sensor detection. In a conventional use of the network, devices sleep more than 90% of the time. Scheme is: devices are sleeping, then wake up, transmit/receive data, then at last go to sleep again (see §III.C.1). Only coordinators stay awake for buffering network messages of sleeping terminals.

#### C. Data link layer

*Data link* layer has two objectives: medium access control and detecting/correcting transmission errors. 802.15.4 also proposes two link level topologies.

*1) Topologies:* 802.15.4 proposes two topologies: peer-to-peer and star.

- Peer-to-peer topology makes possible direct data transfers between FFD. Establishing peer-to-peer links is only possible if devices are in common radio range and use the same radio channel. Medium access is done by using CSMA/CA protocol without RTS/CTS mechanism (§III.C.2). Peer-to-peer topology makes *ad-hoc networks* [8] possible.

- Star topology is possible when a coordinator is present on the channel. The coordinator is a FFD which manages medium access of "slave" nodes. Coordinator's "slave" node can be either a FFD or a RFD. All data transfers are going through the coordinator, making doze mode possible – messages are buffered by the coordinator. This network topology allows high energy saving thanks to an optimal sleeping period distribution between embedded devices. Star coordinator sends beacons frames on demand or periodically; periodically beacons frames make *beacon tracking* possible, a special mode which consists of a RFD in a slept waiting for the next beacon frame, then wakes up, asks the star coordinator for pending data, transmits/receives, and goes to sleep again.

*2) Medium access methods:* IEEE 802.15.4 proposes two medium access methods:

- A method based on CSMA/CA contention method,
- A *contention free method* which allows bandwidth reservation.

The contention method can be used in either peer-to-peer or star topologies but contention free method can be used only in star networks. To use contention free method, star coordinator has to attribute *Guaranteed Time Slots* (GTS) to its nodes, others nodes cannot transmit during these allocated periods. Thanks to contention free medium access, medium reservation is possible and some Quality of Service (QoS) properties like *bandwidth reservation* or *latency guaranties* can be applied. This contention free access is only possible in a *star beaconed network*. Beacon frame contains information which indicates the structure of inter-beacons period named superframe. IEEE 802.15.4 superframe has the following structure:

- At first, star coordinator sends a beacon frame which indicates durations of superframe and sleeping interval, data pending, network and coordinator addresses, cryptographic properties, size of the *Contention Access Period* (CAP) and size of the optional *Contention Free Period* (CFP),
- Coordinator and nodes send/receive frames in the CAP using CSMA/CA protocol (GTS requests are done in the CAP),
- Coordinator or nodes use GTS in the optional CFP,
- Before next beacon frame, the optional sleeping period can be imposed by the star coordinator for energy saving. All network nodes go in *doze mode* (see §III.B) and local timer will wake up each device just before next beacon.

*D. ZigBee: higher layers*

ZigBee is a standard of the ZigBee Alliance [9] which uses IEEE 802.15.4 standard for physical and data-link layers. ZigBee proposes an original protocol suite for 3 to 7 protocol stack levels [10]. With its network layer, ZigBee extends network range and can obtain wide zones, using mesh architecture for example with cluster-tree topology [11]. Thousands of devices can be linked by this way. With its profile system, ZigBee provides a high-level of compatibility between manufacturers. Profiles concept was created for IrDA, and has been widely approved with Bluetooth systems [12].

*E. IEEE 802.15.4 in our application*

According to what was mentioned above, IEEE 802.15.4 is an interesting technology for our application. Topologies adapted to our problem, low power, data encryption, sufficient baudrate and, of course, restricted mobility is possible.

*1) Topologies:* Topology possibilities of 802.15.4 are adapted to our needs: a robot will be equipped with a star network, robot command and control device will be the star coordinator (FFD) and sensors/actuators will be simple network nodes (RFD). Robots can also communicate with other robots via peer-to-peer links using ZigBee mesh topology.

*2) Mobility:* In this application, mobility between different robots is required but will be limited inside the robot: sensors/actuators will not move within the robot or only with a limited mobility, for example a robot arm sensor. Cooperating robots will be coordinated, so mobility is also limited for peer-to-peer network links. Thus network has certain topology stability. 802.15.4 supports "limited mobility" (Access and association delays, time to live and route validity periods)

*3) Energy saving:* Embedded characteristic of the robot makes this point crucial. Other IEEE network technologies like 802.11 (WiFi), 802.15.1 (Bluetooth) are not advanced enough in this respect.

*4) Baudrate and latency:* The application does not need high baudrate but messages must pass through the network with guarantee on delivery time (latency). Network carries crucial sensors and actuators data (obstacle or shock sensor, "stop" motor order, etc.) which can not tolerate delays and must be forwarded in time. Several delay guarantee levels must be defined over the network but medium access without delay guaranty also has to be preserved because all nodes do not need time guarantees. Moreover, the application generates a little traffic (few kilo-octets per second) but it has a very consistent shape. Sensors raw messages are small (generally few octets) and temporal needs are so strict that data cannot be buffered in order to make full network frames. Because of that point, efficient data are very small and frame headers must also be as small as possible. Efficient baudrate will be much reduced as well. The network technology must be the simplest on that point and protocol must be light. 802.15.4 has been conceived in order to present simplest protocol that can be implemented on small 8-bit microcontrollers, that is again an advantage for the application.

*F. Limits of IEEE 802.15.4 in our application*

802.15.4 is an interesting wireless technology but, unfortunately, it is not perfect. For example, in a star network present standard cannot certify *deterministic medium access*. As described in §III.C.2, it is possible to reserve GTS but *GTS requests frames* are sent in the CAP, where frames may generate collisions. This gap is embarrassing, it is impossible to guaranty transmission delays. Even worse: CSMA/CA protocol used in the CAP is based on random message delaying to avoid collisions, which is not adapted for real-time applications.

802.15.4 has another gap: GTS cannot be used in peer-to-peer links. Again, it is not possible to reserve bandwidth, guarantee transmission delay is not possible. These gaps put a stop to QoS exchanges over the network.

The next section presents our proposition on 802.15.4 to eliminate these two disadvantages thanks to a medium access method modification.

## IV. NEW MEDIUM ACCESS MEDIUM PROPOSED

Works presented in this section aim to solve 802.15.4 gaps presented in last section: possibility to make determinist *GTS requests* i.e. without collisions and to avoid collisions in peer-to-peer network links.

### A. GTS requests without collisions

802.15.4 standard makes possible nodes to request GTS to the star coordinator in the CAP, so these requests may cause collisions. To solve this problem, we propose here to reverse the direction of the request. For some important nodes, star coordinator can ask the associated node for possible GTS allocation, rather than keep node doing its GTS request; requests can be done by the coordinator in *a polling*. Coordinator must have *pre acquired knowledge* of its nodes and their needs. In our application this assumption is realistic. This new method is an improvement which permits to guarantee a maximum delay for GTS allocation. Without-priority nodes can still make GTS requests in the CAP and vital nodes can also accelerate their requests by making requests in the CAP.

*1) The Network association requests and GTS requests:* In the life of an IEEE 802.15.4 network, there are two critical temporal phases: network association and GTS requests. In present standard, network association i.e. coordinator association is achieved by a passive listening of the radio medium, searching for beacons frames. All channels must be scanned to find a suitable coordinator; each channel listening period must be calculated according to the beacon transmitting frequency, which must be declared in the node configuration [13]. Once network association is completed, a network node can transmit/receive data, goes in doze mode or request for GTS. These two phases of an IEEE 802.15.4 network life (network association and GTS Requests) are very close to each others if we are searching for temporal guarantees: it may be interesting to obtain a definite response to these requests. Moreover, a definite response is the best to consider energy saving.

*2) Pre acquired knowledge of the neighborhood nodes:* Star coordinator must have pre acquired knowledge of the neighborhood nodes and also the needs of the sensors for making a perfect polling. For our application, this assumption is not restricting because it is the command unit of the robot which organizes the network. This central entity must know all the sensors it has to take care of them. Pre acquired knowledge can be a list of MAC addresses, with a precise description of this node's needs: minimal and maximum baudrate, maximum latency, etc. This knowledge base can be downloaded in the coordinator's memory or created by automatic learning. Controlling first network association can be done by pushing manually synchronization buttons like in DECT systems.

*3) Polling optimization:* Once pre acquired knowledge primitives are implemented, polling has to be optimized: simple linear polling, scheduling with priorities, etc. Several algorithms were developed to solve these classic problems. Each application has its adequate algorithm. Each network star can have its own scheduling.

### B. Peer-to-peer links without collisions

In the previous section, we saw how to solve the deterministic medium access gap of IEEE 802.15.4 standard for GTS requests in a star network topology. In this section, we propose a solution to avoid collisions in a multiple peer-to-peer links area. 802.15.4 standard does not propose a GTS mechanism for peer-to-peer links. In today standard, FFDs can communicate outside the star network with peer-to-peer links using CSMA/CA protocol but because of CSMA/CA employment, present standard guarantees neither baudrate nor transmission delays.

*1) GTS between star coordinators:* Our solution uses the concept of star's GTS. In a star network, each node is associated to a coordinator; this coordinator decides of the medium temporal distribution between nodes. In peer-to-peer links, there is no coordinator; All nodes which are capable of communicating directly are FFD. In this case, it is possible for two FFD to communicate by using the star mode rather than peer-to-peer mode. If a coordinator A wants to send a message to a coordinator B, A must be previously associated to B and requests for a GTS of B's star. In the next superframe, A can send its message to B without collision. GTS temporal position is sent in B's beacon so A must listen to that beacon. Consequently, it is necessary for each coordinator to listen to the next beacon after its GTS request. If a third coordinator C is in the neighborhood, it may not use the medium while B is sending its beacon. Consequently, it is necessary for all network nodes to keep medium unused while beacons messages are being sent. A hard synchronization is required for our temporal organization of the network.

*2) An equal temporal repartition of the beacons:* As a starting point, our solution needs each coordinator to listen to every neighbor's beacons. At first, the hypothesis that all coordinators can listen to all others coordinators is mandatory; in that case, there must be no hidden coordinator. For a robotic application, this hypothesis is not so restrictive. However this simplification must be temporary because our contribution could be used again for another application, for example in a wider area. The hidden coordinator problem can be solved by adding a supercoordinator to our network (for a centralized network solution) which identifies hidden beacons or by creating synchronization and beacon relaying messages between coordinators (for a decentralized network solution).

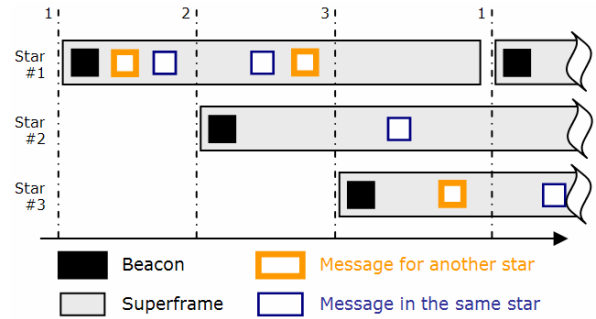

Figure 2. Temporal repartition for 3 stars

The IEEE 802.15.5 task group proposes to avoid beacon collisions of neighbor WPANs to keep independent traffics. It is done by dynamic beacon alignment [14] which consists in hidden coordinator's beacons retransmission.

By using a wide synchronization over the network, simple nodes must also be synchronized: nodes mustn't send frames during beacon transmitting periods. It is necessary to impose reserved timeslots for beacons and reserved timeslots for nodes. The question is to know how beacons reserved timeslots must be positioned to ensure the best repartition of beacons in the superframe and avoid GTS freezing. This problem may appear if there are too many coordinators in the network area: medium will be saturated by beacons; then the first coordinator can allocate all timeslots for its nodes and medium will be immediately full. A first solution consists in an equal time distribution of beacons, as shown in figure 2.

*3) Traffic optimization - coordinator's GTS requests in beacons:* As shown in figure 2, medium temporal distribution is quite condensed. Useless messages have to be limited to optimize bandwidth. In 802.15.4 standard, beacon message size varies with the number of nodes data pending and number of GTS. In a low loaded star, beacon messages are empty. Because beacons are listened to by all coordinators, it may be interesting to use these messages to transport useful information for all coordinators: neighbor coordinators list, received power report, errors report, etc. GTS requests between coordinators can be done via a two beacons request/reply rather than the present standard method, to limit bandwidth loss.

*4) Optimization - variable transmission powers:* Figure 2 shows that various star internal administration messages (GTS requests, acknowledgment, etc.) use baudrate and are not useful for other stars. Another optimization for a coordinator consists in allocating GTS to other coordinators GTS after a preliminary study of received power signal of the beacon or the node message: if a first coordinator allocates some GTS to a node that its messages are always received with a low power by a second coordinator, this coordinator can also allocate a GTS in this timeslot, asking its node to transmit in this time slot with a minimal transmitting power. Two coordinators may allocate the same time-slot for two faraway RFD without making collisions. Medium occupation could be denser as shown in figure 3.

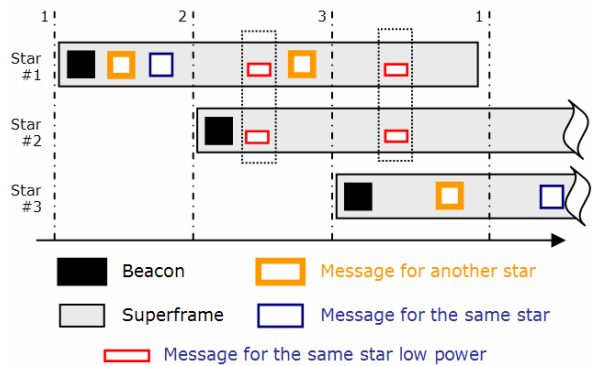

Figure 3. Optimized medium use

## V. CONCLUSION

The work deals with an improvement of actual 802.15.4 standard by adding some QoS features. The proposition consists in a full deterministic medium access method, keeping compatibility with official version of the standard. Two main features are realized: an efficient polling of the nodes by the star coordinator and a temporal deterministic distribution of peer-to-peer messages in a mesh topology. We are validating this protocol to prove its efficiency by adding our contribution in existing 802.15.4 NS2 model [15]. In the same time, we are working on a prototype based on a Freescale development kit to implement our modifications on a couple of real devices. A concrete testing application is also being developed in a robotic prototype [16] in coordination with the Protocols & Networks Research Team of LIMOS Laboratory, University of Clermont Ferrand, France. Future works are numerous: full validation of the network, prototype testing, improvement of the protocol, solving the hidden coordinator problem.

ACKNOWLEDGMENT

These research works are deal with the industrial collaboration of Cyril Zarader, Freescale Semiconductor, Toulouse, France and LIMOS Laboratory, University of Clermont Ferrand, France.